\documentclass[prl,twocolumn,superscriptaddress,showpacs]{revtex4}

\usepackage[dvips]{graphicx}

\renewcommand{\vec}[1]{\mathbf{#1}}

\newcommand{\eref}[1]{(\ref{#1})}

\renewcommand{\(}{\left(}
\renewcommand{\)}{\right)}

\begin{document}
\title{Spectrum of weak magnetohydrodynamic turbulence} 
\author{Stanislav Boldyrev}
\author{Jean Carlos Perez}
\affiliation{Department of Physics, University of Wisconsin-Madison,
  1150 University Ave, Madison, WI 53706, USA}  
\date{\today}

\begin{abstract}
Turbulence of magnetohydrodynamic waves in nature and in the laboratory is generally cross-helical or non-balanced, in that the energies of Afv\'en waves moving in opposite directions along the guide magnetic field are unequal. Based on high-resolution numerical simulations it is proposed that such turbulence spontaneously generates a condensate of the residual energy $E_v-E_b$ at small field-parallel wave numbers. As a result, the energy spectra  of counter-propagating Alfv\'en waves are not scale-invariant. In the limit of infinite Reynolds number, the universality is asymptotically restored at large wave numbers, and both spectra attain the scaling $E(k)\propto k_{\perp}^{-2}$. The generation of condensate is apparently related to the breakdown of mirror symmetry in non-balanced turbulence. 
\end{abstract}

\maketitle

\emph{Introduction.}---Magnetohydrodynamic (MHD) turbulence naturally occurs in a variety of plasmas, ranging from the interstellar medium to the solar wind to laboratory fusion devices. When the compressibility effects can be neglected, the MHD equations take especially simple form in the so-called Els\"asser variables, 
\begin{equation}
  \(\frac{\partial}{\partial t}\mp{\bf v}_A\cdot\nabla\){\bf 
  z}^\pm+\left({\bf z}^\mp\cdot\nabla\right){\bf z}^\pm = -\nabla P+{\bf f}^{\pm},
  \label{mhd-elsasser}
\end{equation}
where ${\bf z}^\pm={\bf 
v}\pm {\bf b}$, ${\bf v}$ is the fluctuating plasma velocity, ${\bf b}$ is
the fluctuating magnetic field normalized by $\sqrt{4 \pi \rho_0}$,
${\bf v}_A={\bf B}_0/\sqrt{4\pi \rho_0}$ is the contribution of the 
uniform magnetic field ${\bf B}_0$, $P=(p/\rho_0+b^2/2)$ includes the
plasma pressure $p$ and the magnetic pressure, $\rho_0$ is the
constant plasma density, ${\bf f}^{\pm}$ represents the mechanisms 
driving turbulence, and small dissipation due to viscosity 
and resistivity is neglected. In the absence of dissipation, both energies $E^+=\langle |z^+|^2 \rangle$ and $E^-=\langle |z^-|^2 \rangle$ are conserved, which is equivalent to the conservation of total energy and cross-helicity~\cite{biskamp}. 

The linear terms, $({\bf v}_A\cdot\nabla) {\bf z}^\pm$, describe advection of Alfv\'en wave packets along the guide field, while the nonlinear interaction terms, $\left({\bf z}^\mp\cdot\nabla\right){\bf z}^\pm$, are responsible for energy redistribution over scales. Depending on the driving force, turbulence exhibits either weak or strong regime in a certain range of scales. Denote $b_{\lambda}$ the rms magnetic fluctuations at the field-perpendicular scale $\lambda \propto 1/k_{\perp}$, and assume that the typical field-parallel wave vector of such fluctuations is~$k_\|$. Then the turbulence is weak when the linear terms dominate, $k_\| v_A\gg k_{\perp} b_{\lambda}$,  and it is strong otherwise. 
 
Weak MHD turbulence may be present in laboratory devices and in the solar wind, as indicated by  energy spectra somewhat steeper that the Kolmogorov one~\citep{biskamp}. Moreover, for a general driving force weak turbulence regime precedes the development of strong turbulence, in a sense that as the scale of fluctuations decreases the turbulence eventually becomes strong.

When the nonlinear interaction is absent, the solution of (\ref{mhd-elsasser}) is an ensemble of shear-Alfv\'en and pseudo-Alfv\'en waves propagating along the guide field~$B_0$ with the velocities~$\pm {\bf v}_A$. The small nonlinear terms then can be taken into account perturbatively, and the spectrum of turbulence can be derived using the general methods of the theory of weak turbulence; for a review see~\cite{zakharov,newell}.
  
Spectra of MHD turbulence were first studied by Iroshnikov~\citep{iroshnikov63} and Kraichnan~\citep{kraichnan}. Those early works realized the role of the guide field in mediating the turbulent cascade, however, they assumed the small scale fluctuations to be isotropic. Over the years, the assumption of anisotropy proved to be incorrect~\citep[e.g.,][]{biskamp,Shebalin}. Anisotropic spectra of weak MHD turbulence were addressed by Ng \& Bhattacharjee~\citep{ng-bhattacharjee} and Goldreich \& Sridhar~\citep{goldreich97} based on dimensional arguments, and a comprehensive analytic framework was developed by Galtier et al~\cite{galtier}.  The latter theory derives the kinetic equations for evolution of the spectral energies $e^{\pm}({\bf k})=\langle |{\bf z}^{\pm}(\bf k)|^2 \rangle $, and has the following main results. 

First, the spectral energies are transferred in the direction of large $k_\perp$, and the universal regime of weak turbulence is established at $k_{\perp}\gg k_\|$. In the  universal regime, the dynamics of shear-Alfv\'en modes decouple from the dynamics of pseudo-Alfv\'en modes, which are passively advected by the shear-Alfv\'en ones. In what follows we shall consider only shear-Alfv\'en waves and keep the same notation $e^{\pm}({\bf k})$ for their energies. It is also customary to use the phase-space-volume compensated spectra defined as  $E^{\pm}({\bf k}) \propto e^{\pm}({\bf k})k_{\perp}$. 

Second, the predicted spectra of weak MHD turbulence are not unique, but form a one-parameter family,  $E^{\pm}(k_{\perp})\propto k_{\perp}^{-2\pm\alpha}$, with $-1<\alpha <1$. The solutions with $\alpha \neq 0$ correspond to unequal fluxes of the $E^{\pm}(k)$-energies over scales; we denote these fluxes $\epsilon^+$ and $\epsilon^-$. It can be further demonstrated that in this case the amplitudes of the energy spectra are different, e.g., $E^+> E^-$ if $\epsilon^+>\epsilon^-$. 
The spectral parameter $\alpha$ is uniquely defined once the ratio $\epsilon^+/\epsilon^-$ is specified. In the balanced case the energy spectrum is $E^{\pm}(k_{\perp})\propto k_{\perp}^{-2}$ \citep{ng-bhattacharjee,goldreich97,galtier,perez-boldyrev}. 

MHD turbulence in nature and in the laboratory is often driven by localized sources (solar wind, antennae, localized instabilities, etc.) and therefore it is generally non-balanced. There is however a more fundamental reason for imbalance in MHD turbulence. Numerical simulations demonstrate that even when MHD turbulence is balanced overall, it gets spontaneously non-balanced locally, that is, it consists of domains of positive and negative imbalance \citep{matthaeus08,perez09,boldyrev09}. This domain structure is apparently related to the conservation of cross-helicity in ideal MHD, and non-balanced turbulence is also called ``cross-helical.'' Non-balanced MHD  turbulence has recently attracted considerable interest~\citep{lithwick03,lithwick07,chandran08,perez09,boldyrev09,beresnyak08}.   

In the present work we study non-balanced MHD turbulence in a series of high resolution numerical simulations. The results reveal puzzling contradictions with the theory. While in the balanced case the numerics confirm the analytic prediction $E(k_{\perp})\propto k_{\perp}^{-2}$ \citep[cf][]{ng,perez-boldyrev}, in a general non-balanced case simulations disagree with the theory.  When we drive turbulence with unequal rates $\epsilon^{\pm}$, the resulting spectra $E^{\pm}$ turn out to be not defined by the ratio $\epsilon^+/\epsilon^-$. Rather, they depend on the Reynolds number and approach $k_{\perp}^{-2}$ at large $k_{\perp}$ as the Reynolds number increases. 

To resolve this contradiction, we propose that driven weak MHD turbulence generates the  residual energy condensate $\langle {\bf z}^+({\bf k})\cdot {\bf z}^{-}(-{\bf k})\rangle = v^2-b^2 \neq 0$ at $k_\|=0$. This condensate has been assumed to be zero in the standard derivation~\citep{ng-bhattacharjee,goldreich97,galtier}, 
therefore, its presence in our model requires an explanation. The Alfv\'en wave fluctuations obey ${\bf v}= \pm {\bf b}$, in which case the residual energy vanishes. However, at $k_\|=0$ fluctuations are not waves, and the  Alfv\'enic condition should not be necessarily satisfied.   
We further propose that the generation of the condensate is a consequence of the breakdown of the mirror symmetry in non-balanced turbulence. \\

\emph{Kinetic equations for weak MHD turbulence.}---
In this section we discuss the predictions of the standard model for weak MHD turbulence developed in \citep{galtier,ng-bhattacharjee,goldreich97,lithwick03}. The kinetic equations for the evolution of the shear-Alfv\'en energies, derived by Galtier et al~\citep{galtier}, have the form: 
\begin{eqnarray}
\partial_t e^{\pm}({\bf k})=\int M_{k,pq}e^{\mp}({\bf q})\left[e^{\pm}({\bf p})-e^{\pm}({\bf k})\right]\delta(q_\|)d_{k,pq},
\label{galtier-eq}
\end{eqnarray}
where we use the short-hand notation $M_{k,pq}=(\pi/v_A)({\bf k}_{\perp}\times {\bf q}_{\perp})^2({\bf k}_{\perp}\cdot {\bf p}_{\perp})^2/(k_{\perp}^2p_{\perp}^2q_{\perp}^2)$ and $d_{k,pq}=\delta({\bf k}-{\bf p}-{\bf q})d^3p\,d^3q$. The derivation assumes that in the zeroth-order approximation, only the correlation functions $e^{\pm}({\bf k})$ are non-zero. As shown in \cite{galtier}, the system (\ref{galtier-eq}) has a degeneracy: the right hand side integrals vanish for  any solutions of the form  
\begin{eqnarray}
e^{\pm}({\bf k})=g^{\pm}(k_\|)k_{\perp}^{-3\pm \alpha},
\label{spectra}
\end{eqnarray} 
with arbitrary functions $g^{\pm}(k_\|)$ and $-1<\alpha<1$. The degeneracy is removed by matching these solutions with the boundary conditions, that is, forcing and dissipation, \citep[e.g.,][]{lithwick03}. To match with the forcing, one notes that different energy spectra correspond to different energy fluxes, $\epsilon^{\pm}$ supplied by the large-scale forcing, such that $\alpha$ is uniquely found if the ratio $\epsilon^+/\epsilon^-$ is specified. One can show that the solution with the steeper spectrum corresponds to the larger energy flux~\cite{galtier}.  

The large-scale boundary conditions fix the slopes of the energy spectra, but do not fix their amplitudes. To fully remove the degeneracy, 
one further argues that at the dissipation scale the balance should be restored, that is,  $e^+(k)$ should converge to $e^-(k)$. This ``pinning'' effect was first pointed out in  \cite{grappin83}, and it physics was discussed in greater detail in \cite{galtier,lithwick03,chandran08}. 
%It is physically understood, if one notes that the $e^+(k)$ and $e^-(k)$ energy spectra cannot intersect in the inertial interval, this would contradict the universality of the turbulence. The alignment of velocity and magnetic fluctuations, preserved by the nonlinear terms, can be broken only by the dissipation. 
The pinning effect is indeed observed in our simulations presented below.  

According to the above picture, if the rates of energy supply are fixed, then the {\em slopes}  of the energy spectra $e^{\pm}(k)$ are fixed as well. If the dissipation scale is now changed, the {\em amplitudes} of the spectra should change as to maintain the specified slopes, and to make them converge at the dissipation scale. This conclusion, although consistent with equations (\ref{galtier-eq}), seems to be at odds with the common intuition about turbulent systems, which suggests that small-scale dissipation should not significantly affect the large-scale fields subject to the same large-scale driving. This seeming contradiction motivated our interest in the problem. \\

{\em The numerical method and the results.}---The universal properties of
MHD turbulence with a strong guide field are accurately described by
neglecting the field-parallel components of the fluctuating fields,
associated with the pseudo-Alfv\'en
mode~\cite{galtier-chandran,perez-boldyrev}.  
%This mode plays a passive role in the dynamics of both weak and strong turbulence. 
By setting
$\vec z_\|^\pm = 0$ in equation \eref{mhd-elsasser} we obtain the
closed system of equations
\begin{eqnarray}
  \(\frac{\partial}{\partial t}\mp\vec v_A\cdot\nabla_\|\)\vec
  z^\pm+\left(\vec z^\mp\cdot\nabla_\perp\right)\vec z^\pm =
  -\nabla_\perp P\nonumber\\ +{\vec f}_{\perp}^\pm+\nu\nabla^2\vec z^\pm,
  \label{rmhd-elsasser}
\end{eqnarray}
in which dissipation terms have been added, and we assume that viscosity is
equal to resistivity.  This set of equations is known as the Reduced
MHD model  originally developed for tokamak plasmas
\cite{kadomtsev,strauss}, and often used in numerical simulations of different regimes of MHD turbulence. Depending on the spectral properties of the driving force, this system can describe either weak or strong MHD turbulence~\citep{galtier-chandran,perez-boldyrev}. 

We employ a fully dealiased Fourier
pseudo-spectral method to solve equations \eref{rmhd-elsasser} with a
strong guide field ($v_A/v_{rms}\sim 5$) in a rectangular periodic
box, with field-perpendicular cross section $L_\perp^2=(2\pi)^2$ and
field-parallel box size $L_\|=5L_{\perp}$. The choice of a rectangular box, as
discussed in~\cite{perez-boldyrev}, allows for correct description of 
long-wavelength and low-frequency fluctuations.

The $z^+$ and $z^-$ waves are driven independently by Gaussian random forces ${\vec f}_{\perp}^+$ and ${\vec f}_{\perp}^-$, with the variances $\sigma^{\pm}=\langle (\vec f_{\perp}^{\pm})^2\rangle$. The imbalance is measured by the parameter $\gamma=(\sigma^+-\sigma^-)/(\sigma^+ + \sigma^-)$.   
To ensure that the turbulence is weak, the forces have a broad $k_\|$ spectrum. They are applied in Fourier space at wave-numbers $1 \leq k_{\perp} \leq 2$ and $(2\pi/L_\|) 
\leq k_\|\leq 16 (2\pi/L_\|)$. The Fourier coefficients inside that range 
are independent Gaussian random numbers with the amplitudes chosen so that the
resulting rms velocity fluctuations are of order unity.  The
individual random values are refreshed independently for each mode on
average every $\tau=0.05~L_\perp/v_{rms}$.  We define 
the Reynolds number as $Re=(L_\perp/2\pi)v_{rms}/\nu$. A typical run covers about~$50$ to~$100$ crossing times at the largest scale.  
\begin{figure}
  \includegraphics[width=0.48\textwidth]{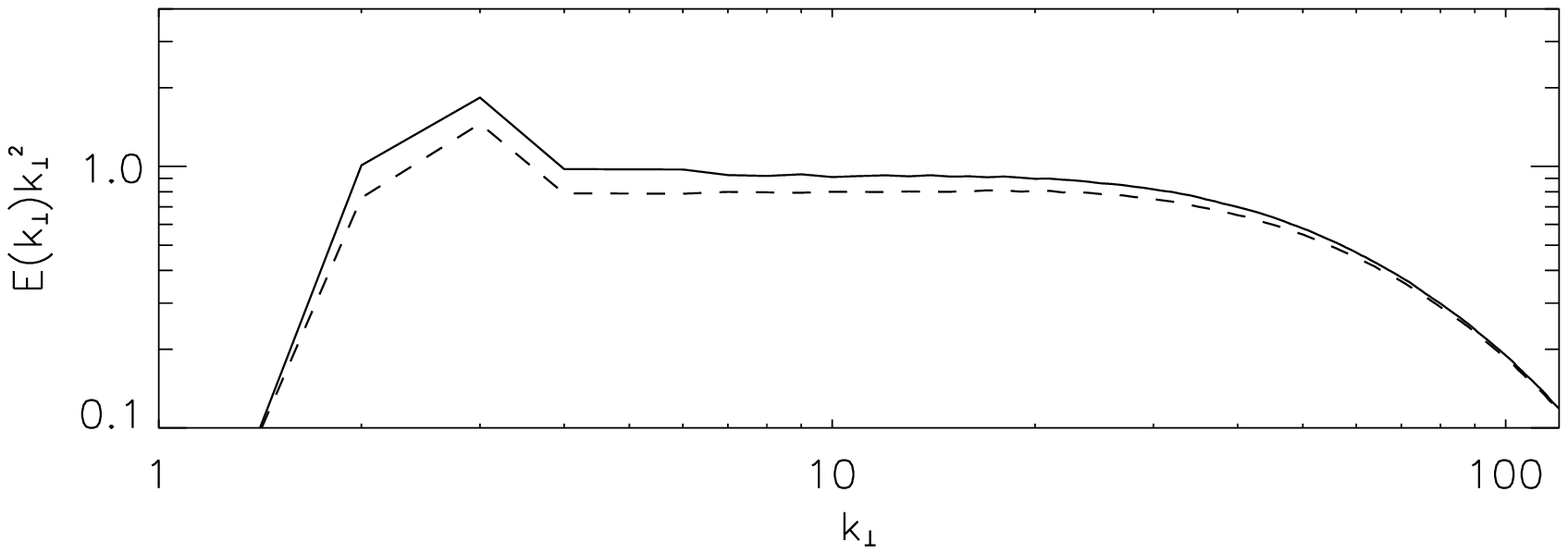}
  \includegraphics[width=0.48\textwidth]{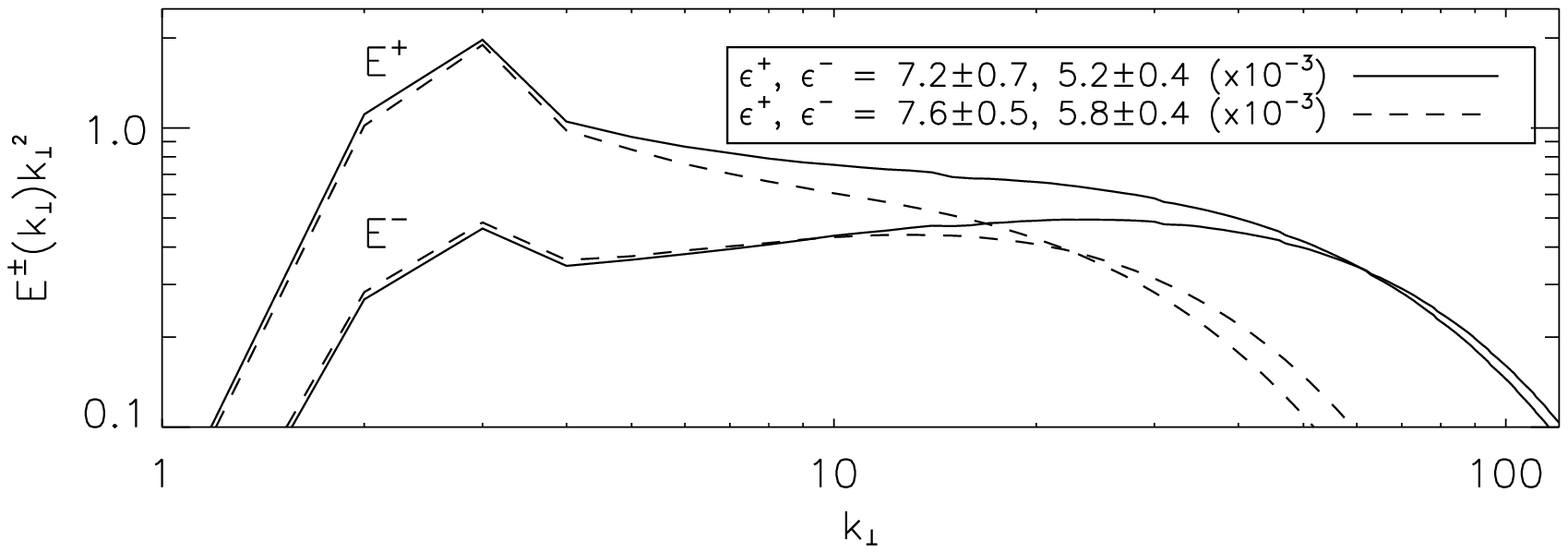}  
  \caption{Upper panel: The spectra of balanced weak MHD turbulence. The solid line is $E^+(k_{\perp})$, the dashed line is $E^-(k_{\perp})$;  $Re=6000$, resolution $1024^2 \times 256$ points. Lower panel: The spectra of non-balanced weak MHD turbulence, with the imbalance parameter~$\gamma=0.17$. The solid lines denote $E^+(k_{\perp})$ and $E^-(k_{\perp})$,    $Re=4500$, resolution $1024^2 \times 256$ points. The dashed lines show the same fields for $Re=2000$ and resolution $512^2\times 256$ points. The insert shows the corresponding energy dissipation rates.}
  \label{pinning}
\end{figure}

As the force renovation time is much shorter than the inverse Alfv\'en frequencies of all the excited modes, the forcing supplies energies at controlled rates, $\epsilon^{\pm}=\frac{1}{2}\sigma^{\pm}$. According to the solution of (\ref{galtier-eq}), in this case the $+$ and $-$ energy slopes should be independent of the dissipation.  The results of our numerical simulations are presented in Fig.~(\ref{pinning}). They demonstrate that the spectra are pinned at the dissipation scale.  However, the amplitudes of the spectra at large scales are not sensitive to the dissipation. As a result, the spectral slopes change with the Reynolds number, as to gradually approach the balanced spectrum $E(k_{\perp})\propto k_{\perp}^{-2}$ at large~$k_{\perp}$. These numerical findings agree with the physical expectation that large-scale fields are determined solely by the large-scale forcing and are independent of the small-scale dissipation. They however contradict the standard model~(\ref{galtier-eq}). In what follows we propose a resolution for this inconsistency.\\

{\em A model for non-balanced weak MHD turbulence.}---To derive the model equations, we propose that non-balanced MHD turbulence leads to the generation of a non-zero average, 
\begin{eqnarray}
%\langle {\bf  z}^+({\bf k})\cdot{\bf z}^-(-{\bf k})\rangle = e^0(k_{\perp})\Delta(k_\|),
\langle {\bf  z}^+({\bf k})\cdot{\bf z}^-({\bf k}')\rangle = \delta ({\bf k}+{\bf k}')e^0(k_{\perp})\Delta(k_\|),
\label{condensate} 
\end{eqnarray}
where $\Delta(k_\|)$ is concentrated at $k_\|=0$. 
If the Elsasser fields ${\bf z}^+$ and ${\bf z}^-$ corresponded to Alfv\'en waves, such an average would be zero, since Alfv\'enic fluctuations satisfy ${\bf v}=\pm{\bf b}$. However, fluctuations at $k_\|=0$ are not waves ($\omega=k_\|v_A=0$), and the average (\ref{condensate}) may not vanish.  The presence of the condensate (\ref{condensate}) means that the magnetic and kinetic energies are not in equipartition at $k_\|=0$. 

\begin{figure}
  \includegraphics[width=0.48\textwidth]{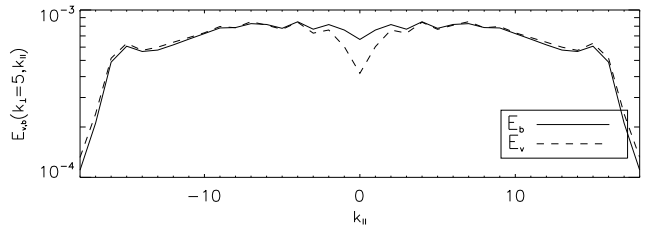}  
  \caption{The field-parallel spectra of the magnetic energy (solid line) and the kinetic energy (dashed line) at the field-perpendicular wave number~$k_{\perp}=5$; $Re=2500$, resolution $512^2\times 256$ points.}
  \label{evb_kpar}
\end{figure}

Physically, non-balanced MHD turbulence is not mirror-invariant, as it possesses the non-zero cross-helicity, $H^C=\int ({\bf v}\cdot {\bf b})d^3x\neq 0$. Non-mirror-invariant turbulence can generate large-scale magnetic fields that are not in equipartition with the velocity field. This happens due to conservation of magnetic helicity in ideal MHD, which tends to cascade toward large scales in a turbulent state. We propose that such a process is preserved in a driven weak MHD turbulence, although in a peculiar fashion --  it leads to generation of  condensate (\ref{condensate}) in the vicinity of $k_\|=0$, where the magnetic energy exceeds the kinetic energy. Such a condensate is indeed observed in our numerics, see Fig.~(\ref{evb_kpar}).

The dynamics of the condensate are not described by the weak turbulence theory, however, they may be addressed using certain closure assumptions. It is worth pointing out that the same limitation holds for the original system (\ref{galtier-eq}). As follows from the presence of the $\delta(q_\|)$-function in the integrand, only the $e^{\mp}(q_\|=0)$ modes are responsible for the energy transfer. However, if one applies  Eq.~(\ref{galtier-eq}) to these modes themselves, one encounters an inconsistency. The weak turbulence approximation is valid when the inverse time of nonlinear interaction is much smaller than the wave frequency. The nonlinear interaction described by the right hand side of (\ref{galtier-eq}) does not vanish for $k_\|=0$, while the linear frequency of the corresponding Alfv\'en waves, $\omega=k_\|v_A$, vanishes. Therefore, as noted in~\cite{galtier}, an additional assumption of smoothness of the functions $g^{\pm}(k_\|)$ at $k_\|=0$ was essential for deriving the spectra~(\ref{spectra}). 

We postpone the discussion of self-consistent condensate equations for future communications. Here we would like to demonstrate the effect that the presence of condensate provides on the turbulence spectra. As we have argued, the condensate is expected to alter the cascade dynamics in the non-balanced case. Consider the case where the imbalance is weak, that is~$\gamma \ll 1$.  In this case we expect that the condensate is weak as well. 
%$e^0({\bf k})\ll e^{\pm}({\bf k})$. 
We  derive the equations for the energies $e^{\pm}({\bf k})$, by proceeding along the lines of weak turbulence derivation: we expand the MHD equations (\ref{mhd-elsasser}) up to the second order in the nonlinear terms, and split the forth-order correlators into the second-order ones according to the Gaussian rule.  To the first order in $e^0(k_{\perp})$, the resulting equations have the form:
\begin{eqnarray} 
\partial_t e^{\pm}({\bf k})=\int M_{k,pq}e^{\mp}({\bf q})\left[e^{\pm}({\bf p})-e^{\pm}({\bf k})\right]\delta(q_\|)d_{k,pq}+ \nonumber \\
+\tilde {\Delta}(k_\|) \int R_{k,pq}\left[e^{\pm}({k_\perp})e^0({q_\perp})+ e^{\pm}(q_\perp)e^0(k_\perp)\right] d^\perp_{k,pq},
\label{bp-eq}
\end{eqnarray}
where $\tilde \Delta(k_\|)={\rm Re}\Delta(k_\|)$, $R_{k,pq}=(\pi/v_A)({\bf k}_{\perp}\times {\bf q}_{\perp})^2({\bf k}_{\perp}\cdot {\bf p}_{\perp})({\bf k}_{\perp}\cdot {\bf q}_{\perp})/(k_{\perp}^2p_{\perp}^2q_{\perp}^2)$, and $d^\perp_{k,pq}=\delta({\bf k}_\perp-{\bf p}_\perp-{\bf q}_\perp)d^2p_\perp\,d^2q_\perp$.  The first term in (\ref{bp-eq}) coincides with the equation (\ref{galtier-eq}), while the second term describes the interaction with the condensate. It can be directly verified that each of the integrals in (\ref{bp-eq}) conserves the Elsasser energies $E^{\pm}=\int e^{\pm}({\bf k})d^3k$. 

For the stationary solution, each of the integrals in (\ref{bp-eq}) should vanish  independently. Equating the first integral to zero does not allow one to find the spectra uniquely, rather, it leads to the one-parameter family of solutions~(\ref{spectra}). Consider the second integral that describes the interaction of~$e^{\pm}$ with the condensate. By employing the standard methods of the weak turbulence theory, one can demonstrate that the power-law solution nullifying the first part of the second integral is unique, $e^0(k_{\perp})\propto k_{\perp}^{-3}$.   
Analogously, the second  part of the second integral is zero if $e^{\pm}(k_{\perp}) \propto k_{\perp}^{-3}$. We conclude that the presence of the condensate lifts the degeneracy of the solutions: the only possible stationary power-law spectra of weak MHD turbulence are  $e^{\pm}(k_{\perp})\propto k_{\perp}^{-3}$. Although we do not have the equation for the condensate, the above result allows us to predict that in order to preserve the scale invariance the condensate should have the scaling $e^0(k_\perp)\propto k_\perp^{-3}$.  \\

{\em Conclusions.}---Based on our numerical simulations and analytic consideration, we propose that weak MHD turbulence spontaneously generates a condensate of the residual energy $E_v-E_b$ at small $k_\|$. We argue that the condensate is a consequence of mirror-symmetry breakdown in non-balanced turbulence. When the turbulence is balanced, the energy spectra are   $E^{\pm}(k_{\perp})\propto k_{\perp}^{-2}$, in agreement with the analytic prediction of  \citep{ng-bhattacharjee,goldreich97,galtier}. In the balanced case the evolution of $E^{\pm}$ fields is not affected by the condensate. In the non-balanced case the interaction with the condensate becomes essential, and we propose that no universal power-law spectra exist in an inertial interval of limited extent. Both spectra $E^{\pm}(k_{\perp})$ have the large-scale amplitudes fully specified by the external forcing, and they converge at the dissipation  scale. As the dissipation scale decreases, the spectral scalings (but not necessary amplitudes) approach each other at large~$k_{\perp}$.  As a result, the universal scaling $k_{\perp}^{-2}$ is recovered for both spectra $E^{\pm}(k_{\perp})$ asymptotically at $k_{\perp}\to \infty$.

\acknowledgments This work was supported by the U.S. DoE Junior Faculty Award under Grant
No.~DE-FG02-07ER54932, and by the NSF Center for Magnetic Self-Organization in Laboratory and Astrophysical Plasmas at the UW-Madison. High Performance Computing resources were
provided by the Texas Advanced Computing Center (TACC) at the
University of Texas at Austin under the NSF-Teragrid Project
TG-PHY080013N.
%URL:http://www.tacc.utexas.edu   

%\bibliographystyle{apsrev}
% Here the bibliography
%\bibliography{references}

\end{document}